\documentstyle[sprocl,epsfig,psfig,aps]{revtex}

\input{psfig}

\def\ra{\rightarrow}

\def\be{\begin{equation}}
\def\ee{\end{equation}}
\def\bea{\begin{eqnarray}}
\def\eea{\end{eqnarray}}

\newcommand{\newc}{\newcommand} 
\newc{\lra}{\leftrightarrow} 
\newc{\beq}{\begin{equation}} 
\newc{\eeq}{\end{equation}} 
\newc{\barr}{\begin{eqnarray}} 
\newc{\earr}{\end{eqnarray}} 

\begin{document} 
\title{SUSY Dark Matter in the Universe- Theoretical  Direct Detection Rates}

\author{J. D. VERGADOS} 

\address{Theoretical Physics Section, University of Ioannina, GR-45110, 
Greece\\E-mail:Vergados@cc.uoi.gr} 

\maketitle\abstracts{
Exotic dark matter together with the vacuum energy or  cosmological constant 
seem to dominate in the Universe. An even higher
density of such matter seems to be gravitationally trapped in our Galaxy.
Thus its direct detection 
is central to particle physics and cosmology. Current fashionable supersymmetric
models provide a natural dark matter candidate which is the lightest
supersymmetric particle (LSP). Such models combined with fairly well 
understood physics like
the quark substructure of the nucleon and the nuclear structure (form factor 
and/or spin response function), permit the evaluation of
the event rate for LSP-nucleus elastic scattering. The thus obtained event rates
are, however, very low or even undetectable.
 So it is imperative to exploit the modulation effect, i.e. the dependence of
the event rate on  the earth's annual motion.  Also it is useful to consider
the directional rate, i.e its dependence on the direction of the recoiling
nucleus. In this paper we study
such a modulation effect both in non directional and directional experiments.
We calculate both the differential and the total rates using both isothermal, 
symmetric as well
as only axially asymmetric, and non isothermal, due to caustic rings, velocity
distributions. We find that in the symmetric case 
the modulation amplitude is small. The same is true for the case of caustic
 rings.  The inclusion of asymmetry, with a realistic enhanced velocity 
dispersion  in the galactocentric direction, yields an enhanced
modulation effect, especially in directional experiments.}
\section{Introduction}
In recent years the consideration of exotic dark matter has become necessary
in order to close the Universe \cite{Jungm}. Furthermore in
in order to understand the large scale structure of the universe 
it has become necessary to consider matter
made up of particles which were 
non-relativistic at the time of freeze out. This is  the cold dark 
matter component (CDM). The COBE data ~\cite{COBE} suggest that CDM
is at least $60\%$ ~\cite {GAW}. On the other hand during the last few years
evidence has appeared from two different teams,
the High-z Supernova Search Team \cite {HSST} and the
Supernova Cosmology Project  ~\cite {SPF} $^,$~\cite {SCP} 
 which suggests that the Universe may be dominated by 
the  cosmological constant $\Lambda$.
As a matter of fact recent data the situation can be adequately
described by  a baryonic component $\Omega_B=0.1$ along with the exotic 
components $\Omega _{CDM}= 0.3$ and $\Omega _{\Lambda}= 0.6$
(see next section for the definitions).
In another analysis Turner \cite {Turner} gives 
$\Omega_{m}= \Omega _{CDM}+ \Omega _B=0.4$.
Since the non exotic component cannot exceed $40\%$ of the CDM 
~\cite{Jungm},~\cite {Benne}, there is room for the exotic WIMP's 
(Weakly  Interacting Massive Particles).
  In fact the DAMA experiment ~\cite {BERNA2} 
has claimed the observation of one signal in direct detection of a WIMP, which
with better statistics has subsequently been interpreted as a modulation signal
~\cite{BERNA1}.

The above developments are in line with particle physics considerations. Thus,
in the currently favored supersymmetric (SUSY)
extensions of the standard model, the most natural WIMP candidate is the LSP,
i.e. the lightest supersymmetric particle. In the most favored scenarios the
LSP can be simply described as a Majorana fermion, a linear 
combination of the neutral components of the gauginos and Higgsinos
\cite{Jungm,ref1,Gomez,ref2}. 

\section{Density Versus Cosmological Constant}

 The evolution of the Universe is governed by the General Theory of
Relativity. The most
  commonly used  model is that of Friedman, which utilizes the Robertson-
Walker metric
\begin{equation}
 (ds)^2=(dt)^2-R^2(t)[\frac{(dr)^2}{1-\kappa r^2}
          +r^2((d \theta)^2+ \sin ^2 \theta (d \phi)^2)]
\label{cosmo.1}
\end{equation}
The resulting Einstein equations are:
\begin{equation}
 R_{\mu \nu}-\frac{1}{2}g_{\mu \nu}R=-8\pi G_{N} 
T_{\mu \nu}+
                                      \Lambda g_{\mu \nu}
\label{kosmo.2}
\end{equation}
where $G_{N}$ is Newton's constant and $\Lambda$ is the
cosmological constant.
The equation for the scale factor $(t)$ becomes:
\begin{equation}
\frac{d^{2}R}{dt^{2}}=-\frac{4\pi}{3}G_{N}(\rho+3p)R=-\frac{4\pi
                   G_{N}\rho}{3}R+\frac{\Lambda}{3}
\label{cosmo.3}
\end{equation}
where $\rho$ is the mass density. Then the energy is
\beq
E=\frac{1}{2}m(\frac{dr}{dt})^{2}-G_{N}\frac{m}{R}(4\pi\rho
R^{3})+\frac{\Lambda}{6}mR^{2}=constant=-\frac{\kappa}{2}m
\label{kosmo.4}
\eeq
 This can be equivalently be written as
\beq
 H^{2}+\frac{\kappa}{R^{2}}=\frac{8\pi}{3}G_{N\rho}+\frac{\Lambda}{3}
\label{cosmo.5a}
\eeq
 where the quantity $H$ is Hubble's constant defined by
\beq
 H=\frac{1}{R}\frac{dR}{dt}
\label{cosmo.5b}
\eeq
Hubble's constant is perhaps the most important parameter of cosmology. In
fact it is not a constant but it changes with time. Its present day value
is given by
\beq
H_0 =( 65\pm 15)~km/s~M_{pc}^{-1}
\label{cosmo5.c}
\eeq
In other words $H_0^{-1}=(1.50\pm).35)\times10^{10}~y$, which is roughly equal
to the age of the Universe. Astrophysicists conventionally write it as
\beq
H_0 =100~h~km/s~M_{pc}^{-1}~~~~,~~~~~0.5<h<0.8
\label{cosmo5.d}
\eeq
 Equations \ref{cosmo.3}-\ref{cosmo.5a} coincide with those of the
Newtonian
  theory with the following two types of forces:
  An attractive force decreasing in absolute value with the scale
  factor (Newton)
  and a repulsive force increasing with the scale factor (Einstein)

\beq
  F=-G_{N} \frac{mM}{R^{2}}~~(Newton)~~,~~
  F=\frac{1}{3}\Lambda mR~~(Einstein)
\label{cosmo.6b}
\eeq
   Historically the cosmological constant was introduced by
   Einstein so that General Relativity yields a stationary
   Universe, i.e. one which satisfies the conditions:
\beq
    \frac{dR}{dt}=0~~~~\frac{d^{2}R}{dt^{2}}=0
\label{cosmo.7}
\eeq
    Indeed for $\kappa>0$, the above equations lead to
    $R=R_{c}=constant$ provided that
\beq
    \frac{1}{3}\Lambda R_{c}-\frac{4\pi}{3} G_{N}\rho R_{c}=0~~,~~
     \frac{1}{3}\Lambda R^{2}_{c}-\frac{4\pi}{3} G_{N}\rho R^{2}_{c}=\kappa
\label{cosmo.8}
\eeq
    These equations have a non trivial solution provided that the density
$\rho$ and the cosmological constant $\Lambda$ are related, i.e.
\beq
    \Lambda=4 \pi G_{N} \rho
\label{cosmo.9}
\eeq
    The radius of the Universe then is given by
\beq
    R_{c}=[\frac{\kappa}{4\pi G_{N}\rho}]^{1/2}
\label{cosmo.10}
\eeq
    Define now
\beq
\Omega_{m}=\frac{\rho}{\rho _{c}}~~,~~
          \Omega_{\Lambda}=\frac{\rho _{v}} {\rho _{c}}~~,~~
    \rho_{v}=\frac{\Lambda}{8 \pi G_{N}}~~("vacuum" density)
\label{cosmo.11}
\eeq
    The critical density is
\beq
    \rho_{c}=1.8\times
    10^{-23}h^{2}\frac{g}{cm^{3}}=10h^{2}\frac{nucleons}{m^{3}}
\label{cosmo.13}
\eeq
    With these definitions Friedman's equation
    $E=-\kappa\frac{m}{2}$
    takes the form
\beq
    \frac{\kappa}{R^{2}}=(\Omega _{m} +\Omega _{\Lambda}-1)H^{2}
\label{cosmo.14}
\eeq
    Thus we distinguish the following special cases:
\beq
    \kappa >0~~~~\Leftrightarrow~~~~\Omega _{m} +\Omega
    _{\Lambda}>1~~~~\Leftrightarrow Closed~curved~Universe
\eeq
\beq
    \kappa =0~~~~\Leftrightarrow~~~~\Omega _{m} +\Omega
    _{\Lambda}=1~~~~\Leftrightarrow Open~ Flat~ Universe
\eeq
\beq
    \kappa <0~~~~\Leftrightarrow~~~~\Omega _{m} +\Omega
    _{\Lambda}<1~~~~\Leftrightarrow Open~ Curved~ Universe
\eeq
   In other words it is the combination of matter and "vacuum" 
   energy, which determines the fate of the our Universe.

    Before concluding this section we remark that the above
    equations do not suffice to yield a solution since the density
    is a function of the scale factor. An equation of state is in
    addition needed, but we are not going to elaborate further.

\section{An Overview of Direct Detection - The Allowed SUSY Parameter Space.}

 Since this particle is expected to be very massive, $m_{\chi} \geq 30 GeV$, and
extremely non relativistic with average kinetic energy $T \leq 100 KeV$,
it can be directly detected ~\cite{JDV96,KVprd} mainly via the recoiling
of a nucleus (A,Z) in the elastic scattering process:
\begin{equation}
\chi \, +\, (A,Z) \, \to \, \chi \,  + \, (A,Z)^* 
\end{equation}
($\chi$ denotes the LSP). In order to compute the event rate one needs
the following ingredients:

1) An effective Lagrangian at the elementary particle 
(quark) level obtained in the framework of supersymmetry as described 
, e.g., in Refs.~\cite{Jungm,ref2}.

2) A procedure in going from the quark to the nucleon level, i.e. a quark 
model for the nucleon. The results depend crucially on the content of the
nucleon in quarks other than u and d. This is particularly true for the scalar
couplings as well as the isoscalar axial coupling ~\cite{Dree}$^-$\cite{Chen}.

3) Compute the relevant nuclear matrix elements \cite{KVdubna,DIVA00}
using as reliable as possible many body nuclear wave functions.
By putting as accurate nuclear physics input as possible, 
one will be able to constrain the SUSY parameters as much as possible.
The situation is a bit simpler in the case of the scalar coupling, in which
case one only needs the nuclear form factor.

Since the obtained rates are very low, one would like to be able to exploit the
modulation of the event rates due to the earth's
revolution around the sun \cite{Verg98,Verg99}$^-$\cite{Verg01}. To this end one
adopts a folding procedure
assuming some distribution~\cite{Jungm,Verg99,Verg01} of velocities for the LSP.
One also would like to know the directional rates, by observing the 
nucleus in a certain direction, which correlate with the motion of the
sun around the center of the galaxy and the motion of the Earth 
\cite {ref1,UKDMC}.

 The calculation of this cross section  has become pretty standard.
 One starts with   
representative input in the restricted SUSY parameter space as described in
the literature~\cite{Gomez,ref2}. 
We will adopt a phenomenogical procedure taking  universal soft 
SUSY breaking terms at $M_{GUT}$, i.e., a 
common mass for all scalar fields $m_0$, a common gaugino mass 
$M_{1/2}$ and a common trilinear scalar coupling $A_0$, which 
we put equal to zero (we will discuss later the influence of 
non-zero $A_0$'s). Our effective theory below $M_{GUT}$ then 
depends on the parameters \cite{Gomez}:
\[
m_0,\ M_{1/2},\ \mu_0,\ \alpha_G,\ M_{GUT},\ h_{t},\ ,\ h_{b},\ ,\ h_{\tau},\ 
\tan\beta~,  
\]
where $\alpha_G=g_G^2/4\pi$ ($g_G$ being the GUT gauge coupling 
constant) and $h_t, h_b, h_\tau $ are respectively the top, bottom and 
tau Yukawa coupling constants at $M_{GUT}$. The values of $\alpha_G$ and 
$M_{GUT}$ are obtained as described in Ref.\cite{Gomez}.
For a specified value of $\tan\beta$ at $M_S$, we determine $h_{t}$ at 
$M_{GUT}$ by fixing the top quark mass at the center of its 
experimental range, $m_t(m_t)= 166 \rm{GeV}$. The value
of  $h_{\tau}$ at $M_{GUT}$ is  fixed by using the running tau lepton
mass at $m_Z$, $m_\tau(m_Z)= 1.746 \rm{GeV}$. 
The value of $h_{b}$ at $M_{GUT}$ used is such that:
\[
m_b(m_Z)_{SM}^{\overline{DR}}=2.90\pm 0.14~{\rm GeV}.
\]
after including the SUSY threshold correction. 
The SUSY parameter space is subject to the
 following constraints:\\
1.) The LSP relic abundance will satisfy the cosmological constrain:
\begin{equation}
0.09 \le \Omega_{LSP} h^2 \le 0.22
\label{eq:in2}
\end{equation}
 2.) The Higgs bound obtained from recent CDF \cite{VALLS}
 and LEP2 \cite{DORMAN},
 i.e. $m_h~>~113~GeV$.\\
3.) We will limit ourselves to LSP-nucleon cross sections for the scalar
coupling, which gives detectable rates
\begin{equation}
4\times 10^{-7}~pb~ \le \sigma^{nucleon}_{scalar} 
\le 2 \times 10^{-5}~pb~
\label{eq:in3}
\end{equation}
 We should remember that the event rate does not depend only
 on the nucleon cross section, but on other parameters also, mainly
 on the LSP mass and the nucleus used in target. 
The condition on the nucleon cross section imposes severe constraints on the
acceptable parameter space. In particular in our model it restricts 
$tan \beta$ to values $tan \beta \simeq 50$. We will not elaborate further
on this point, since it has already appeared \cite{gtalk}. 
\bigskip
\section{Expressions for the Differential Cross Section .} 
\bigskip

 The effective Lagrangian describing the LSP-nucleus cross section can
be cast in the form \cite {JDV96}
 \beq
{\it L}_{eff} = - \frac {G_F}{\sqrt 2} \{({\bar \chi}_1 \gamma^{\lambda}
\gamma_5 \chi_1) J_{\lambda} + ({\bar \chi}_1 
 \chi_1) J\}
 \label{eq:eg 41}
\eeq
where
\beq
  J_{\lambda} =  {\bar N} \gamma_{\lambda} (f^0_V +f^1_V \tau_3
+ f^0_A\gamma_5 + f^1_A\gamma_5 \tau_3)N~~,~~
J = {\bar N} (f^0_s +f^1_s \tau_3) N
 \label{eq:eg.42}
\eeq

We have neglected the uninteresting pseudoscalar and tensor
currents. Note that, due to the Majorana nature of the LSP, 
${\bar \chi_1} \gamma^{\lambda} \chi_1 =0$ (identically).

 With the above ingredients the differential cross section can be cast in the 
form \cite{ref1,Verg98,Verg99}
\begin{equation}
d\sigma (u,\upsilon)= \frac{du}{2 (\mu _r b\upsilon )^2} [(\bar{\Sigma} _{S} 
                   +\bar{\Sigma} _{V}~ \frac{\upsilon^2}{c^2})~F^2(u)
                       +\bar{\Sigma} _{spin} F_{11}(u)]
\label{2.9}
\end{equation}
\begin{equation}
\bar{\Sigma} _{S} = \sigma_0 (\frac{\mu_r(A)}{\mu _r(N)})^2  \,
 \{ A^2 \, [ (f^0_S - f^1_S \frac{A-2 Z}{A})^2 \, ] \simeq \sigma^S_{p,\chi^0}
        A^2 (\frac{\mu_r(A)}{\mu _r(N)})^2 
\label{2.10}
\end{equation}
\begin{equation}
\bar{\Sigma} _{spin}  =  \sigma^{spin}_{p,\chi^0}~\zeta_{spin}~~,~~
\zeta_{spin}= \frac{(\mu_r(A)/\mu _r(N))^2}{3(1+\frac{f^0_A}{f^1_A})^2}S(u)
\label{2.10a}
\end{equation}
\begin{equation}
S(u)=[(\frac{f^0_A}{f^1_A} \Omega_0(0))^2 \frac{F_{00}(u)}{F_{11}(u)}
  +  2\frac{f^0_A}{ f^1_A} \Omega_0(0) \Omega_1(0)
\frac{F_{01}(u)}{F_{11}(u)}+  \Omega_1(0))^2  \, ] 
\label{2.10b2}
\end{equation}
\begin{equation}
\bar{\Sigma} _{V}  =  \sigma^V_{p,\chi^0}~\zeta_V 
\label{2.10c}
\end{equation}
\begin{equation}
\zeta_V =  \frac{(\mu_r(A)/\mu _r(N))^2}{(1+\frac{f^1_V}{f^0_V})^2} A^2 \, 
(1-\frac{f^1_V}{f^0_V}~\frac{A-2 Z}{A})^2 [ (\frac{\upsilon_0} {c})^2  
[ 1  -\frac{1}{(2 \mu _r b)^2} \frac{2\eta +1}{(1+\eta)^2} 
\frac{\langle~2u~ \rangle}{\langle~\upsilon ^2~\rangle}] 
\label{2.10d}
\end{equation}
\\
$\sigma^i_{p,\chi^0}=$ proton cross-section,$i=S,spin,V$ given by:\\
$\sigma^S_{p,\chi^0}= \sigma_0 ~(f^0_S)^2~(\frac{\mu _r(N)}{m_N})^2$ 
 (scalar) , 
(the isovector scalar is negligible, i.e. $\sigma_p^S=\sigma_n^S)$\\
$\sigma^{spin}_{p,\chi^0}=\sigma_0~~3~(f^0_A+f^1_A)^2~(\frac{\mu _r(N)}{m_N})^2$ 
  (spin) ,
$\sigma^{V}_{p,\chi^0}= \sigma_0~(f^0_V+f^1_V)^2~(\frac{\mu _r(N)}{m_N})^2$ 
(vector)   \\
where $m_N$ is the nucleon mass,
 $\eta = m_x/m_N A$, and
 $\mu_r(A)$ is the LSP-nucleus reduced mass,  
 $\mu_r(N)$ is the LSP-nucleon reduced mass and  
\begin{equation}
\sigma_0 = \frac{1}{2\pi} (G_F m_N)^2 \simeq 0.77 \times 10^{-38}cm^2 
\label{2.7} 
\end{equation}
\begin{equation}
Q=Q_{0}u~~, \qquad Q_{0} = \frac{1}{A m_{N} b^2}=4.1\times 10^4A^{-4/3}~KeV 
\label{2.15} 
\end{equation}
where
Q is the energy transfer to the nucleusr,
$F(u)$ is the nuclear form factor and
\begin{equation}
F_{\rho \rho^{\prime}}(u) =  \sum_{\lambda,\kappa}
\frac{\Omega^{(\lambda,\kappa)}_\rho( u)}{\Omega_\rho (0)} \,
\frac{\Omega^{(\lambda,\kappa)}_{\rho^{\prime}}( u)}
{\Omega_{\rho^{\prime}}(0)} 
, \qquad \rho, \rho^{\prime} = 0,1
\label{2.11} 
\end{equation}
are the spin form factors \cite{KVprd} ($\rho , \rho^{'}$ are isospin indices)
normalized to one at $u=0$.
$\Omega_0$ ($\Omega_1$) are the static isoscalar (isovector) spin 
matrix elements.
Note that the quantity S(u) is essentially independent of $u$. So the energy
transfer dependence is contained in the function $F_{11}(u)$. Note also that
$S(u)$ depends on the ratio of the isoscalar to isovector axial current 
couplings. These individual couplings can vary a lot within the SUSY parameter
space.
\begin{table}[t]  
\caption{
The static spin matrix elements for the light nuclei considered 
here. For comparison we also quote the results for the medium heavy nucleus 
$^{73}$Ge \protect\cite{Ress} and the heavy nucleus $^{207}$Pb
 \protect\cite{KVprd}.
}
\label{table.spin}
\begin{center}
\begin{tabular}{lrrrrr}
 &   &  &  &  &    \\
 & $^{19}$F & $^{29}$Si & $^{23}$Na  & $^{73}$Ge & $^{207}$Pb\\
\hline
    &   &  &  &  &    \\
$[\Omega_{0}(0)]^2$         & 2.610   & 0.207  & 0.477  & 1.157    & 0.305\\
$[\Omega_{1}(0)]^2$         & 2.807   & 0.219  & 0.346  & 1.005    & 0.231\\
$\Omega_{0}(0)\Omega_{1}(0)$& 2.707   &-0.213  & 0.406  &-1.078    &-0.266\\
$\mu_{th} $& 2.91   &-0.50  & 2.22  &    &\\
$\mu_{exp}$& 2.62   &-0.56  & 2.22  &    &\\
$\mu_{th}(spin)/ \mu_{exp}(spin)$& 0.91   &0.99  & 0.57  &    &\\
\end{tabular}
\end{center}
\end{table}

\begin{figure}
\epsfig{figure=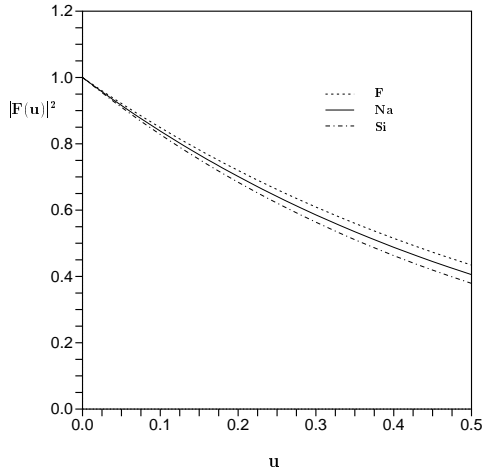,height=3.90in,angle=0}
\vspace*{-2.cm}
\caption{
The energy dependence of the coherent process, i.e. the square
 of the form factor, ($|F(u)|^2$),
for the isotopes $^{19}F$,$^{23}Na$ and $^{29}Si$.
The allowed range of  
$u$ for the above isotopes is $0.011\leq u \leq 0.17$, 
$0.015\leq u \leq 0.30$, and $0.021\leq u \leq 0.50$ respectively. 
This corresponds to energy transfers $8.9\leq Q \leq 140$, 
$9.5\leq Q \leq 190$, and $9.7\leq u \leq 230$ $KeV$ respectively. 
\label{cohff}}
\end{figure}
\begin{figure}
\epsfig{figure=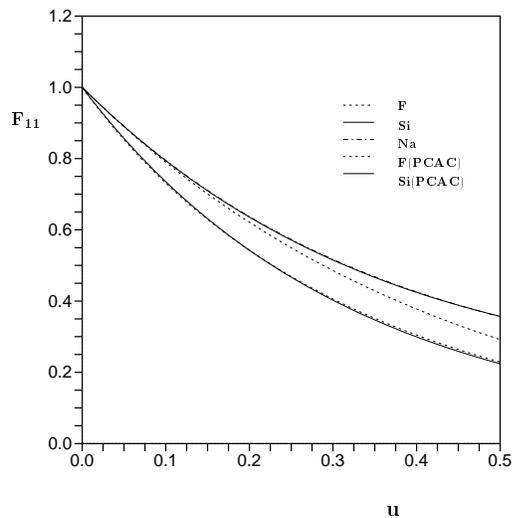,height=3.90in,angle=0}
\vspace*{-2.cm}
\caption{
The energy dependence of the spin contribution (spin response function
 $F_{11}(u)$)
for the isotopes $^{19}F,^{23}Na$ and $^{29}Si$. The allowed range
of energy transfers is the same as in Table \ref{cohff}
In this figure we also plot  
$F_{11}(u)$ when the PCAC effect is  considered. 
\label{spinff}}
\end{figure}

 Their ratio, however, is not changing very much. In fact actual
 calculations \cite{DIVA00} show that $3.0\leq S(0)\leq < 7.5$ for $^{19}F$,
$0.03\leq S(0)\leq 0.2$ for $^{29}Si$ and
$0.4\leq S(0)\leq 1.1$ for $^{23}Na$. The quantity $S(u)$ depends very
 sensitively on nuclear physics via the static spin ME. This is exhibited
in Table (\ref{table.spin}). As we can see from Table \ref{table.spin} the
spin matrix elements are very accurate. This is evident by comparing the
obtained magnetic dipole moments to experiment and noting that the magnetic
 moments, with the exception of $^{23}Na$  are dominated by the spin. From
 the same table we see that $^{19}F$ is
favored from the point of view of the spin matrix element. This advantage 
may be partially lost if the LSP is very heavy, due to the kinematic factor
$\mu_r (A)$, which tends to favor a heavy target.
The energy transfer dependence of the differential cross section for the
coherent mode is given
 by the square of the form factor,
 i.e. $|F(u)|^2$. These form factors
for the isotopes $^{19}F$,$^{23}Na$ and $^{29}Si$ were calculated by
Divari {\it et al} \cite{DIVA00} and are shown in Fig (\ref{cohff}).
The energy transfer dependence of the differential cross section due to 
spin is essentially given by $F_{11}(u)$.  These functions
for the isotopes $^{19}F,^{23}Na$ and $^{29}Si$ were calculated by
Divari {\it et al} \cite{DIVA00} and are shown in Fig (\ref{spinff}).
Note that the energy dependence of the coherent and the spin modes for 
light systems are not very different, especially if the PCAC corrections
on the spin response function are ignored.

\section{Expressions for the Rates.} 
 The non-directional event rate is given by:
\begin{equation}
R=R_{non-dir} =\frac{dN}{dt} =\frac{\rho (0)}{m_{\chi}} \frac{m}{A m_N} 
\sigma (u,\upsilon) | {\boldmath \upsilon}|
\label{2.17} 
\end{equation}
 Where
 $\rho (0) = 0.3 GeV/cm^3$ is the LSP density in our vicinity and 
 m is the detector mass 
The differential non-directional  rate can be written as
\begin{equation}
dR=dR_{non-dir} = \frac{\rho (0)}{m_{\chi}} \frac{m}{A m_N} 
d\sigma (u,\upsilon) | {\boldmath \upsilon}|
\label{2.18}  
\end{equation}
where $d\sigma(u,\upsilon )$ was given above.

 The directional differential rate \cite{ref1},\cite{Verg01} in the
direction $\hat{e}$ is given by :
\begin{equation}
dR_{dir} = \frac{\rho (0)}{m_{\chi}} \frac{m}{A m_N} 
{\boldmath \upsilon}.\hat{e} H({\boldmath \upsilon}.\hat{e})
 ~\frac{1}{2 \pi}~  
d\sigma (u,\upsilon)
\label{2.20}  
\end{equation}
where H the Heaviside step function. The factor of $1/2 \pi$ is 
introduced, since  the differential cross section of the last equation
is the same with that entering the non-directional rate, i.e. after
an integration
over the azimuthal angle around the nuclear momentum has been performed.
In other words, crudely speaking, $1/(2 \pi)$ is the suppression factor we
 expect in the directional rate compared to the usual one. The precise 
suppression factor depends, of course, on the direction of observation.
In spite of their very interesting experimental signatures, we will
not be concerned here with directional rates.
The mean value of the non-directional event rate of Eq. (\ref {2.18}), 
is obtained by convoluting the above expressions with the LSP velocity
distribution $f({\bf \upsilon}, {\boldmath \upsilon}_E)$ 
with respect to the Earth, i.e. is given by:
\beq
\Big<\frac{dR}{du}\Big> =\frac{\rho (0)}{m_{\chi}} 
\frac{m}{A m_N}  
\int f({\bf \upsilon}, {\boldmath \upsilon}_E) 
          | {\boldmath \upsilon}|
                       \frac{d\sigma (u,\upsilon )}{du} d^3 {\boldmath \upsilon} 
\label{3.10} 
\eeq
 The above expression can be more conveniently written as
\beq
\Big<\frac{dR}{du}\Big> =\frac{\rho (0)}{m_{\chi}} \frac{m}{Am_N} \sqrt{\langle
\upsilon^2\rangle } {\langle \frac{d\Sigma}{du}\rangle } 
\label{3.11}  
\eeq
where
\beq
\langle \frac{d\Sigma}{du}\rangle =\int
           \frac{   |{\boldmath \upsilon}|}
{\sqrt{ \langle \upsilon^2 \rangle}} f({\boldmath \upsilon}, 
         {\boldmath \upsilon}_E)
                       \frac{d\sigma (u,\upsilon )}{du} d^3 {\boldmath \upsilon}
\label{3.12}  
\eeq

 After performing the needed integrations over the velocity distribution,
to first order in the Earth's velocity, and over the energy transfer u  the
 last expression takes the form
\beq
R =  \bar{R}~t~
          [1 + h(a,Q_{min})cos{\alpha})] 
\label{3.55a}  
\eeq
where $\alpha$ is the phase of the Earth ($\alpha=0$ around June 2nd)
and  $Q_{min}$ is the energy transfer cutoff imposed by the detector.
In the above expressions $\bar{R}$ is the rate obtained in the conventional 
approach \cite {JDV96} by neglecting the folding with the LSP velocity and the
momentum transfer dependence of the differential cross section, i.e. by
\beq
\bar{R} =\frac{\rho (0)}{m_{\chi}} \frac{m}{Am_N} \sqrt{\langle
v^2\rangle } [\bar{\Sigma}_{S}+ \bar{\Sigma} _{spin} + 
\frac{\langle \upsilon ^2 \rangle}{c^2} \bar{\Sigma} _{V}]
\label{3.39b}  
\eeq
where $\bar{\Sigma} _{i}, i=S,V,spin$ have been defined above, see Eqs
 (\ref {2.10}) - (\ref {2.10c}). It contains all the parameters of the
SUSY models. 
 The modulation is described by the parameter $h$ . Once
the rate is known and the parameters $t$ and $h$, which depend only
on the LSP mass, the nuclear form factor and the velocity distribution 
the nucleon cross section can be extracted and compared to experiment.

The total  directional event rates  can be obtained in a similar fashion by
by integrating Eq. (\ref {2.20}) 
with respect to the velocity as well as the energy transfer u.  We find

\barr
R_{dir}& = &  \bar{R} [(t^0/4 \pi) \, 
            |(1 + h_1(a,Q_{min})cos{\alpha}) {\bf e}~_z.{\bf e}
\nonumber\\  &-& h_2(a,Q_{min})\, 
cos{\alpha} {\bf e}~_y.{\bf e}
                      + h_3(a,Q_{min})\, 
sin{\alpha} {\bf e}~_x.{\bf e}|
\label{4.55}  
\earr
We remind that the z-axis is in the direction of the sun's motion, the y-axis
 is perpendicular to the plane of the galaxy and the x-axis is in the 
galactocentric direction.
 The effect of folding
with LSP velocity on the total rate is taken into account via the quantity
$t^0$, which depends on the LSP mass. All other SUSY parameters have been
 absorbed in $\bar{R}$.
 We see that the modulation of the directional total event rate can
be described in terms of three parameters $h_l$, l=1,2,3. 
 In the special case of $\lambda=0$ we essentially have  one 
parameter, namely $h_1$, since then we have $h_2=0.117$ and $h_3=0.135$.
Given the functions $h_l(a,Q_{min})$ one can plot the the expression in
Eq. (\ref {4.55}) as a function of the phase of the earth $\alpha$. 
\section{The Scalar Contribution- The Role of the Heavy Quarks}
\bigskip

The coherent scattering can be mediated via the the neutral
intermediate Higgs particles (h and H), which survive as physical 
particles. It can also be mediated via s-quarks, via the mixing
of the isodoublet and isosinlet s-quarks of the same charge. In our
model we find that the Higgs contribution becomes dominant and, as a
matter of fact the heavy Higgs H is more important (the Higgs particle
$A$ couples in a pseudoscalar way, which does not lead to coherence).
It is well known that all quark flavors contribute
\cite{Dree}, since the relevant couplings are proportional to
the quark masses.
One encounters in the nucleon not only the usual 
sea quarks ($u {\bar u}, d {\bar d}$ and $s {\bar s}$) but the 
heavier quarks $c,b,t$ which couple to the nucleon via two gluon 
exchange, see e.g. Drees {\it et al}  ~\cite{Dree00} and references
therein.
 
As a result  one obtains an effective scalar Higgs-nucleon
coupling by using  effective quark masses as follows
\begin{center}
$m_u \ra f_u m_N, \ \ m_d \ra f_d m_N. \ \ \  m_s \ra f_s m_N$   
\end{center}
\begin{center}
$m_Q \ra f_Q m_N, \ \ (heavy\ \  quarks \ \ c,b,t)$   
\end{center}
where $m_N$ is the nucleon mass. The isovector contribution is now
negligible. The parameters $f_q,~q=u,d,s$ can be obtained by chiral
symmetry breaking 
terms in relation to phase shift and dispersion analysis.
Following Cheng and Cheng ~\cite{Chen} we obtain:
\begin{center}
$ f_u = 0.021, \quad f_d = 0.037, \quad  f_s = 0.140$ 
\quad  \quad  (model B)   
\end{center}
\begin{center}
$ f_u = 0.023, \quad f_d = 0.034, \quad  f_s = 0.400$ 
\quad  \quad  (model C)   
\end{center}
 We see that in both models the s-quark is dominant.
Then to leading order via quark loops and gluon exchange with the
nucleon one finds:
\begin{center}
\quad $f_Q= 2/27(1-\quad \sum_q f_q)$   
\end{center}
This yields:
\begin{center}
\quad $ f_Q = 0.060$    (model B),   
\quad $ f_Q = 0.040$    (model C)   
\end{center}
 There is a correction to the above parameters coming from loops
involving s-quarks \cite {Dree00} and due to QCD effects. 
 Thus for large $tan \beta$ we find \cite {ref1}:
\begin{center}
\quad $f_{c}=0.060 \times 1.068=0.064,
       f_{t}=0.060 \times 2.048=0.123,
       f_{b}=0.060 \times 1.174=0.070$  \quad (model B)
\end{center}
\begin{center}
\quad $f_{c}=0.040 \times 1.068=0.043,
       f_{t}=0.040 \times 2.048=0.082,
       f_{b}=0.040 \times 1.174=0.047$  \quad (model B)
\end{center}
For a more detailed discussion we refer the reader to 
Refs~\cite{Dree,Dree00}.

\section{Results and Discussion}
\begin{figure}
\epsfig{figure=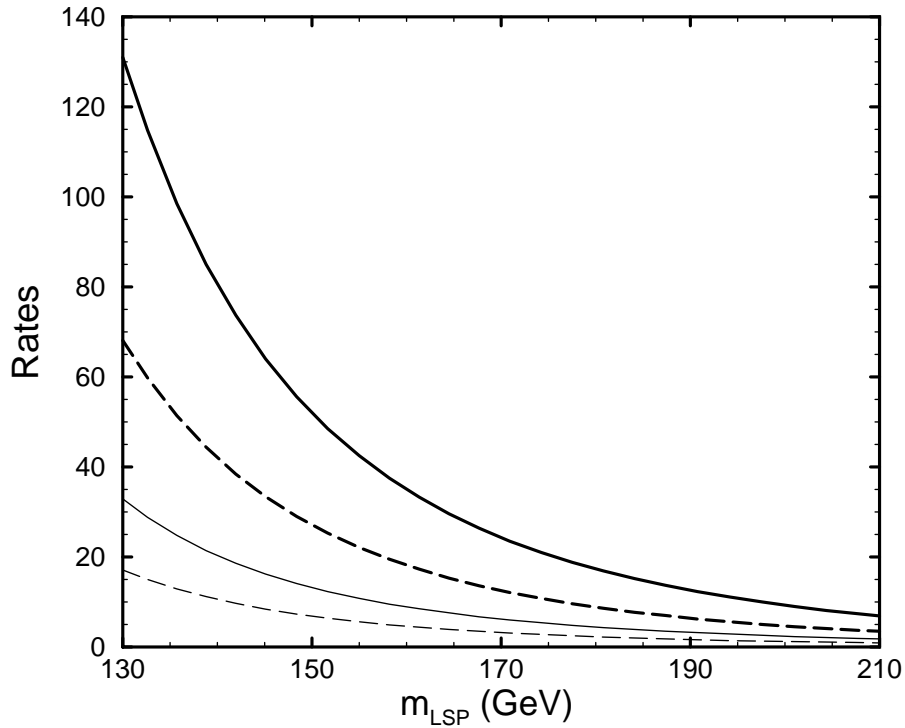,height=3.90in,angle=0}
\caption{The Total detection rate per $(kg-target)yr$ vs the LSP mass
in GeV for a typical solution in our parameter space in the case of 
$^{127}I$ corresponding to  model B (thick line) and Model C (fine 
line). For the definitions see text.
\label{rate}}
\end{figure}

The three basic ingredients of our calculation were 
the input SUSY parameters (see sect. 1), a quark model for the nucleon
(see sect. 3) and the velocity distribution combined with the structure of
 the nuclei involved (see sect. 2). we will focus our attention on the
coherent scattering and present results for the popular target $^{127}I$.
We have utilized two nucleon models indicated by B and C which take into
account the presence of heavy quarks in the nucleon. We also considered
energy cut offs imposed by the detector,  
 by considering two  typical cases $Q_{min}=10,~20$ KeV. The thus obtained
results for the unmodulated total non directional event rates $\bar{R}t$
 in the case
 of the symmetric isothermal model for a typical SUSY parameter choice
\cite{Gomez} are shown in Fig. \ref{rate}. 
Special attention was paid to the the directional rate and its  
modulation due to the annual motion of the earth in the case of isothermal
models. The case of non isothermal models, e.g. caustic rings, is more
complicated \cite{Verg01} and it will not be further discussed here.
 As expected, the parameter $t_0$, which contains the 
effect of the nuclear form factor and the LSP velocity dependence,
decreases as the reduced mass increases.
 
We will focus to the discussion of the directional rates
 described in terms of $t_0$
and  $h_i,~i=1,2,3$ (see Eq.  (\ref{4.55})) 
 and limit ourselves to directions of observation 
close to the coordinate axes. 
 As expected, the parameter $t_0$,
decreases as the reduced mass increases.
The quantity $t^0$ is shown in Fig. (\ref{t0}),  
for three values of the detector energy cutoff
, $Q_{min}=0,10$ and $20~KeV$. Similarly we show the quantity
$h_1$ in Fig. (\ref{h1}). The quanities 
$h_2$ and $h_3$  are shown in Fig (\ref{h23}) for $\lambda=1$.
For $\lambda=0$ they are not shown, since they are 
essentially constant and equal to $0.117$ and $0.135$ respectively. 
\setlength{\unitlength}{1mm}
\begin{figure}
\begin{picture}(150,50)
\put(0,04){\epsfxsize=5cm \epsfbox{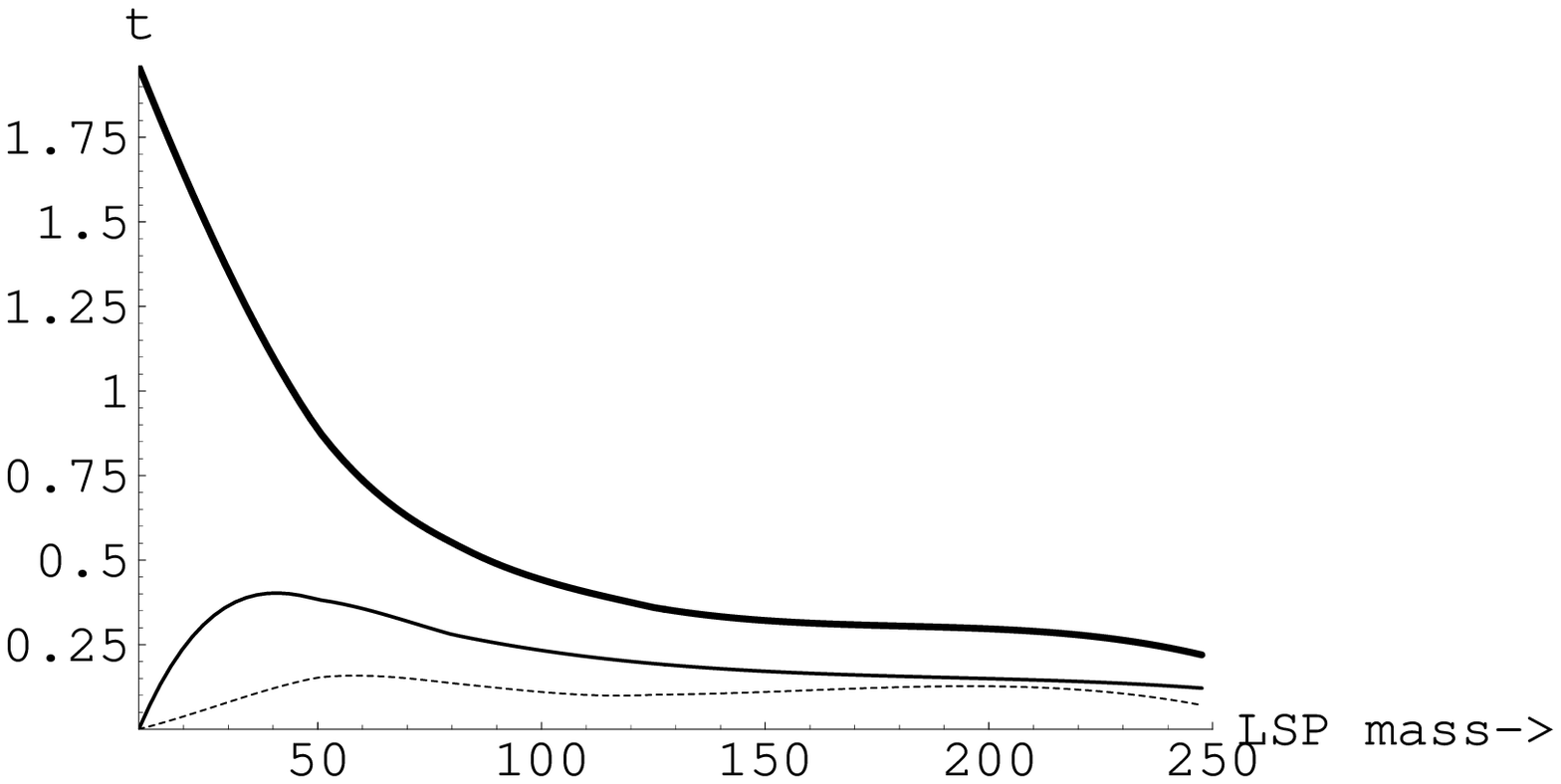}}
\put(8,04) {$\lambda=0$}
\put(60,04){\epsfxsize=5cm \epsfbox{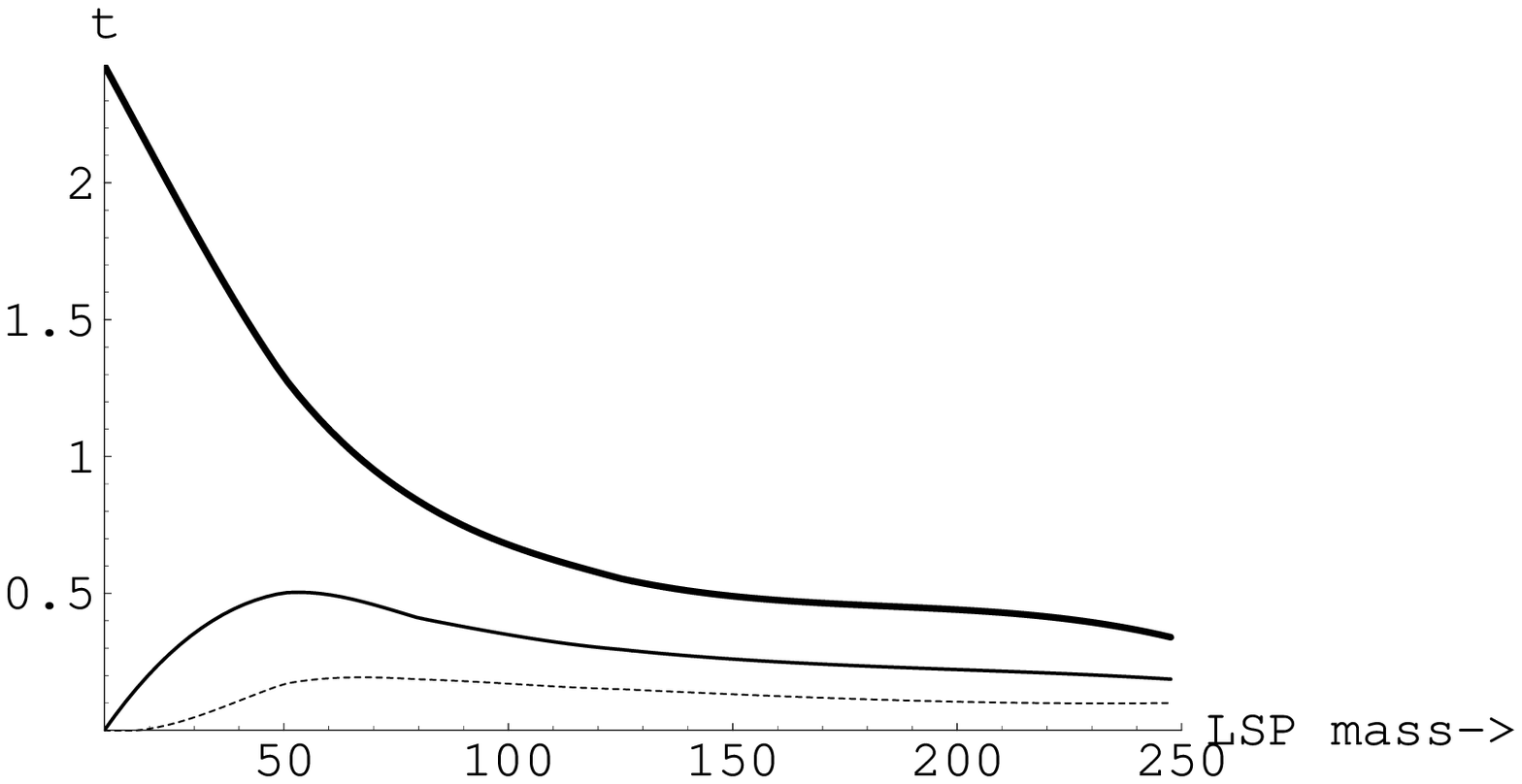}}
\put(85,04){$\lambda=1$}
\end{picture}
\caption[]{
The dependence of the quantity $t^0$ on the LSP mass
 for the symmetric case ($\lambda=0$) as well as for the maximum axial
 asymmetry ($\lambda=1$) in the case of the target $^{127}I$.
For orientation purposes  three  detection cutoff energies are exhibited,
$Q_{min}=0$ (thick solid line),$ Q_{min}=5~keV$ (thin solid line) and
$ Q_{min}=10~keV$ (dahed line).
 As expected $t^0$ decreases as the cutoff energy increases. 
}
\label{t0}
\end{figure}
\setlength{\unitlength}{1mm}
\begin{figure}
\begin{picture}(150,50)
\put(0,04){\epsfxsize=5cm \epsfbox{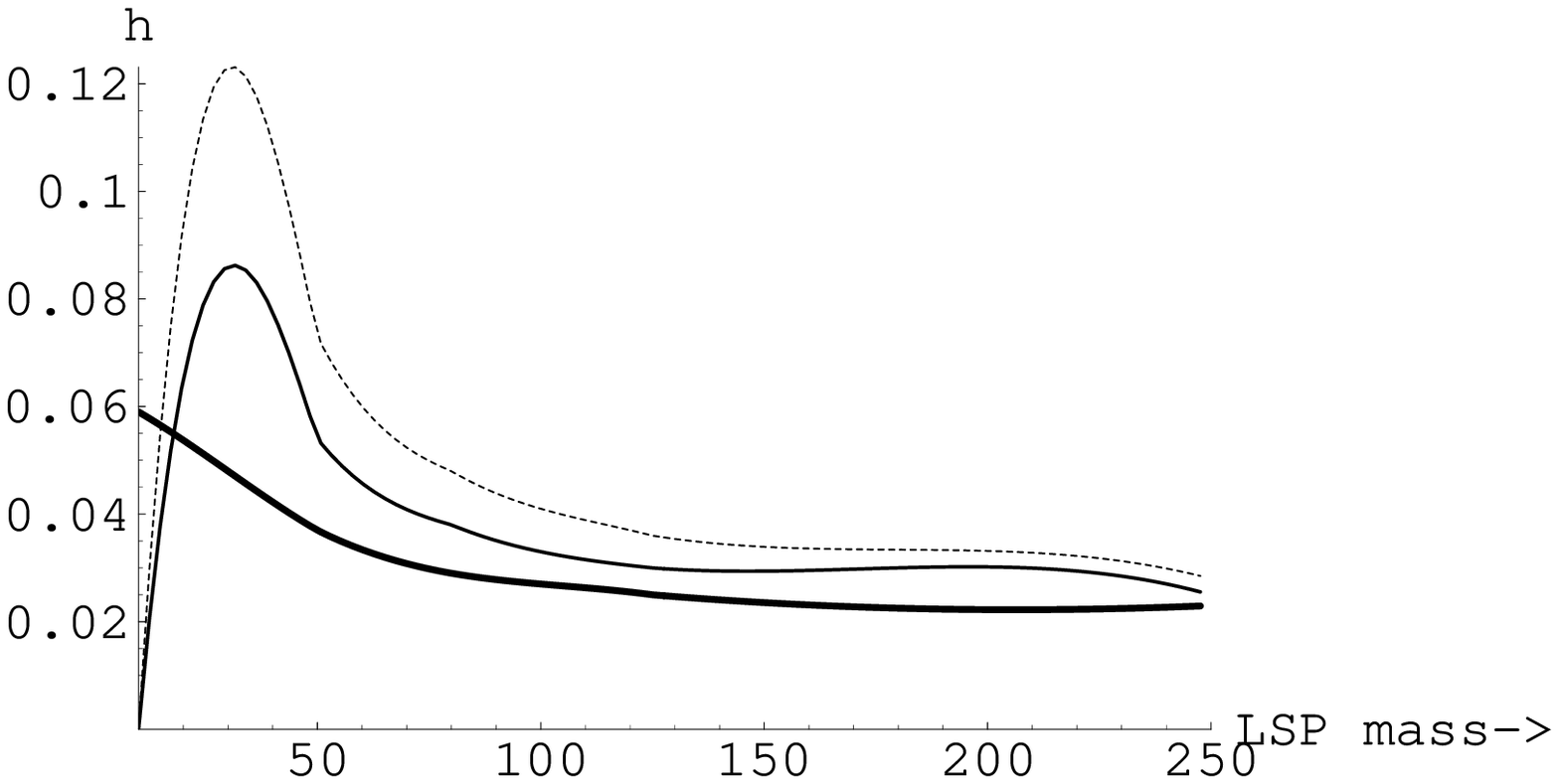}}
\put(8,04) {$\lambda=0$}
\put(60,04){\epsfxsize=5cm \epsfbox{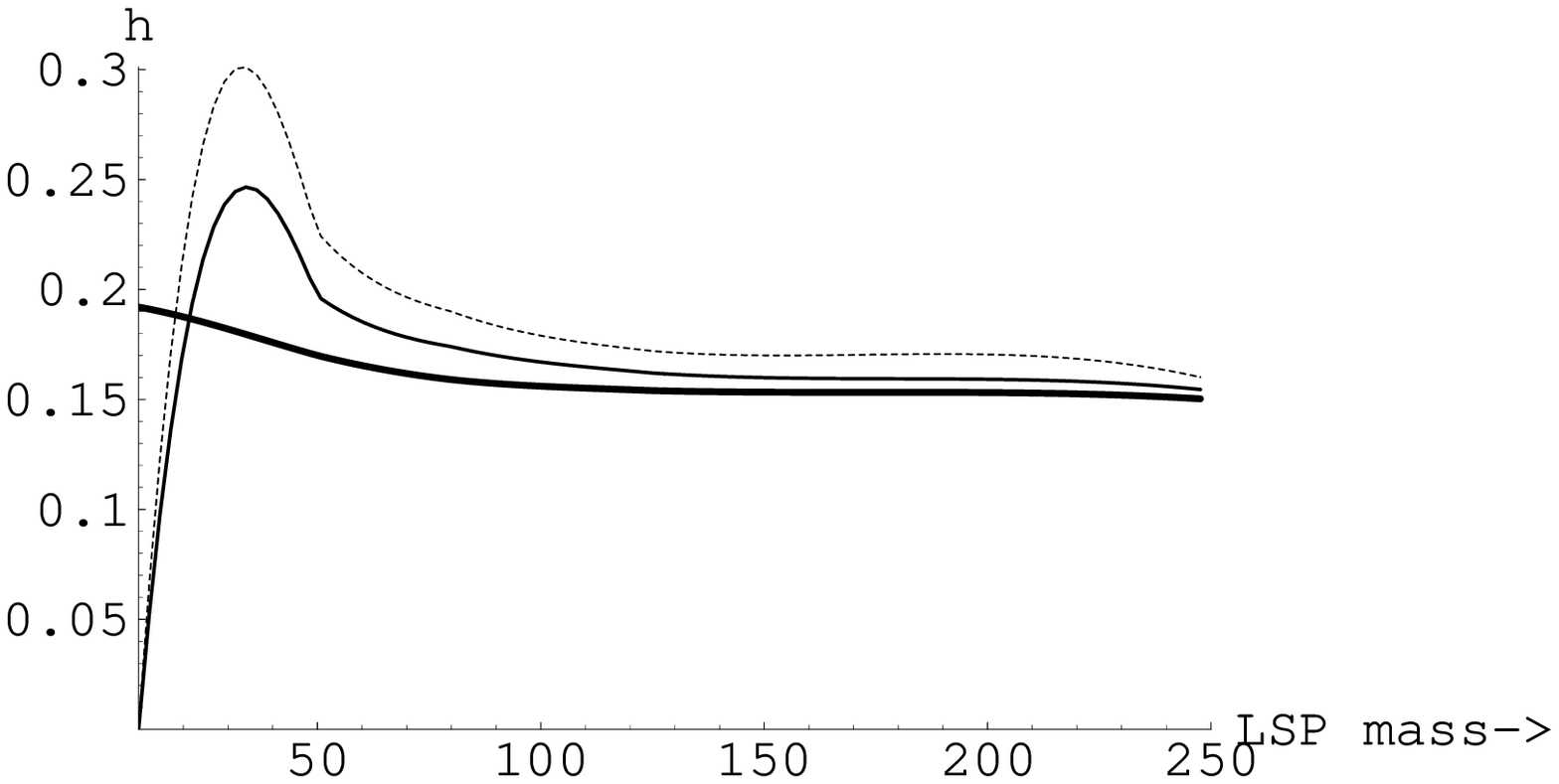}}
\put(85,04){$\lambda=1$}
\end{picture}
\caption[]{
The same as in the previous figure for the modulation amplitude $h_1$.
}
\label{h1}
\end{figure}
\setlength{\unitlength}{1mm}
\begin{figure}
\begin{picture}(150,50)
\put(0,04){\epsfxsize=5cm \epsfbox{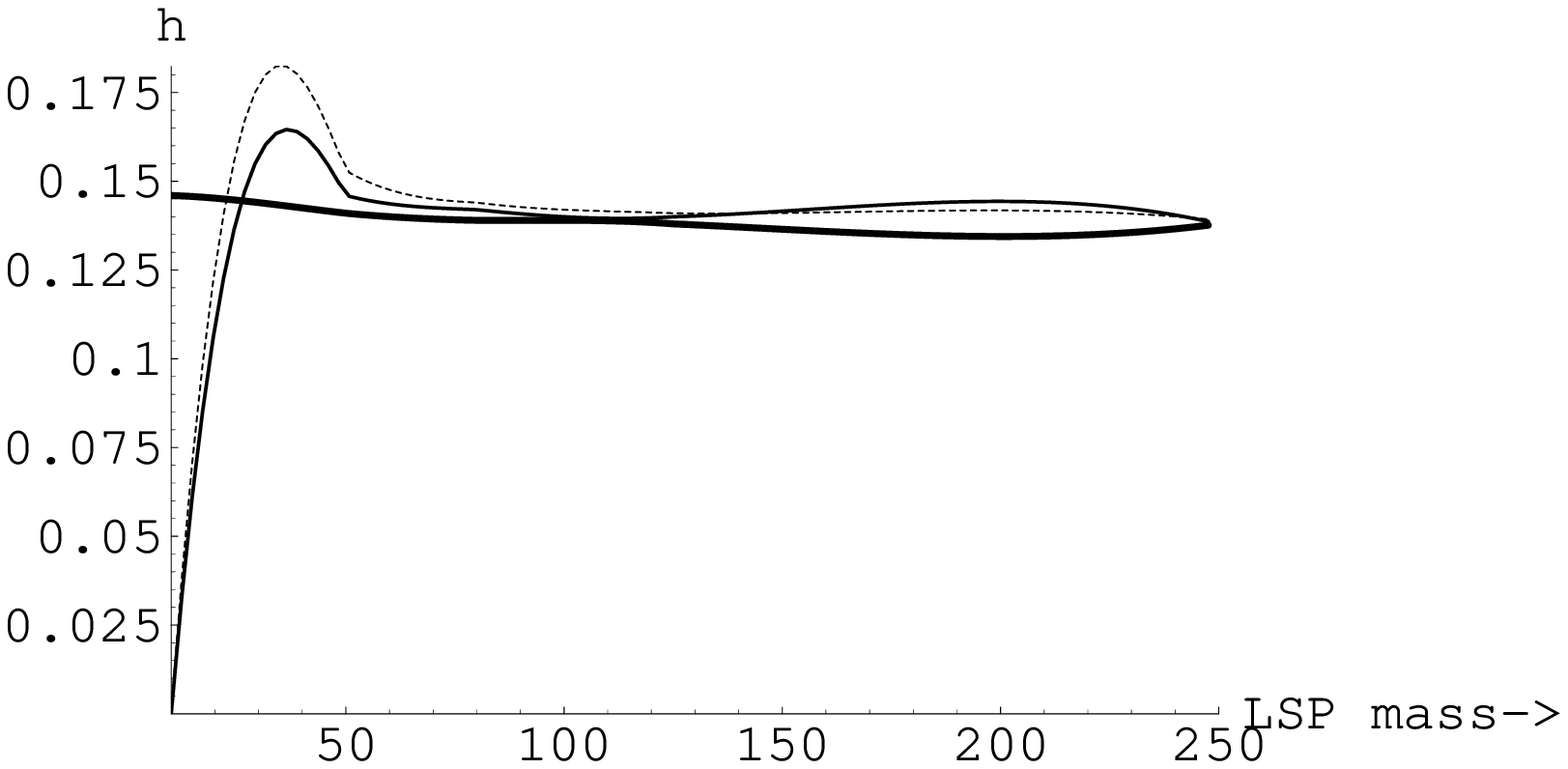}}
\put(6,04) {$h_2$ (for $\lambda=1$)}
\put(60,04){\epsfxsize=5cm \epsfbox{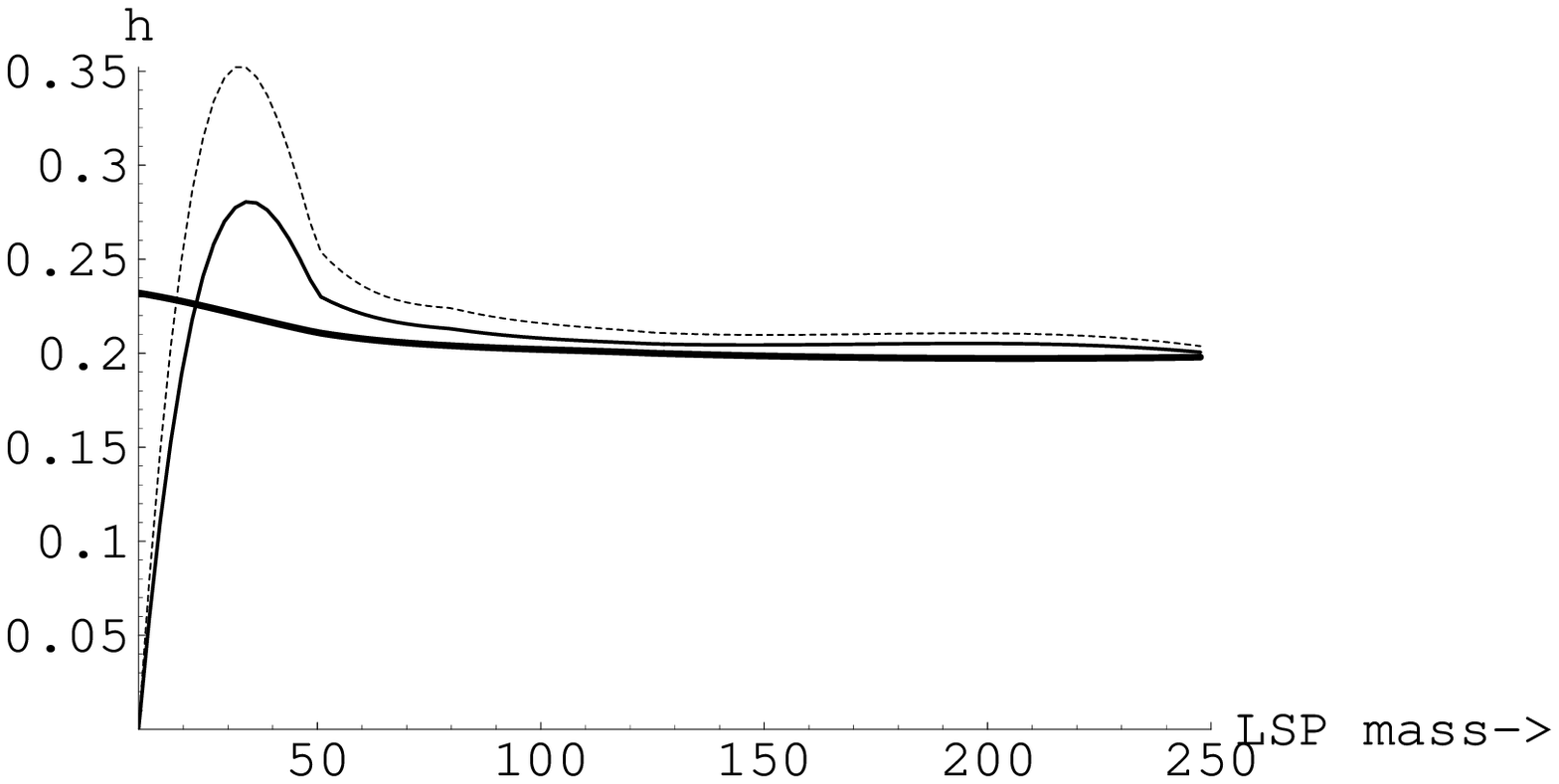}}
\put(75,04){$h_3$ (for $\lambda=1$)}
\end{picture}
\caption[]{
The same as in the previous figure for  the modulation amplitudes $h_2$
and  $h_3$ for $\lambda=1$.
}
\label{h23}
\end{figure}
 As expected, the parameter $t_0$, decreases as the reduced mass increases.
It also decreasres as the cutoff energy $Q_{min}$ increases.
 We notice that $t^0$ is affected little by the presence of asymmetry.
 On the other
hand $h_1,h_2$ and $h_3$ substantially increase in the presence of asymmetry.
Sometimes they increase as the cutoff energy increases (at the expense, of
 course, of the total number of counts. 
For the differential rate the reader is 
referred to our previous work \cite {Verg99,Verg00}.
\section{Conclusions}
In the present paper we have discussed the parameters, which describe
the event rates for direct detection of SUSY dark matter.
Only in a small segmant of the allowed parameter space the rates are above
 the present experimental goals.
 We thus looked for
characteristic experimental signatures for background reduction, i.e.
a) Correlation of the event rates
with the motion of the Earth (modulation effect) and b)
 the directional rates (their correlation  both with the velocity of the sun
 and that of the Earth.)

A typical graph for the total unmodulated rate is shown Fig. \ref{rate}. 
We will concentrate here on the directional rates, described in
terms of the parameters $t_0,h_1,h_2$ and $h_3$.
For simplicity these parameters are given in Figs (\ref{t0})-(\ref{h23})
for directions
of observation close to the three axes $x,y,z$.  We see that the unmodulated 
rate scales by the $cos\theta_s$, with $\theta_s$ (the angle between
the direction of observation and the velocity of the sun).
 The reduction factor, $f_{red}=t_0/(4 \pi~t_0)=\kappa/(2 \pi)$, 
 of the total directional
rate, along the sun's direction of motion, compared to the total non
directional rate depends on the
nuclear parameters, the reduced mass and the asymmetry parameter $\lambda$
\cite{Verg00}.
 We find
that $\kappa$ is around 0.6 (no asymmetry) and around 0.7 (maximum
asymmetry, $\lambda=1.0$), i.e. not very different from the 
naively expected $f_{red}=1/( 2 \pi)$, i.e.  $\kappa=1$. 
The modulation of the directional rate 
increases with the asymmetry parameter $\lambda$ and it also depends
of the direction of observation. For $Q_{min}=0$ it can reach values up 
to $23\%$. Values up to $35\%$ are possible for large values of
 $Q_{min}$, but they occur at the expense of the total
number of counts.

\par
This work was supported by the European Union under the contracts 
RTN No HPRN-CT-2000-00148 and TMR 
No. ERBFMRX--CT96--0090 and $\Pi E N E \Delta~95$ of the Greek 
Secretariat for Research.

\def\ijmp#1#2#3{{ Int. Jour. Mod. Phys. }{\bf #1~}(#2)~#3}
\def\pl#1#2#3{{ Phys. Lett. }{\bf B#1~}(#2)~#3}
\def\zp#1#2#3{{ Z. Phys. }{\bf C#1~}(#2)~#3}
\def\prl#1#2#3{{ Phys. Rev. Lett. }{\bf #1~}(#2)~#3}
\def\rmp#1#2#3{{ Rev. Mod. Phys. }{\bf #1~}(#2)~#3}
\def\prep#1#2#3{{ Phys. Rep. }{\bf #1~}(#2)~#3}
\def\pr#1#2#3{{ Phys. Rev. }{\bf D#1~}(#2)~#3}
\def\np#1#2#3{{ Nucl. Phys. }{\bf B#1~}(#2)~#3}
\def\npps#1#2#3{{ Nucl. Phys. (Proc. Sup.) }{\bf B#1~}(#2)~#3}
\def\mpl#1#2#3{{ Mod. Phys. Lett. }{\bf #1~}(#2)~#3}
\def\arnps#1#2#3{{ Annu. Rev. Nucl. Part. Sci. }{\bf
#1~}(#2)~#3}
\def\sjnp#1#2#3{{ Sov. J. Nucl. Phys. }{\bf #1~}(#2)~#3}
\def\jetp#1#2#3{{ JETP Lett. }{\bf #1~}(#2)~#3}
\def\app#1#2#3{{ Acta Phys. Polon. }{\bf #1~}(#2)~#3}
\def\rnc#1#2#3{{ Riv. Nuovo Cim. }{\bf #1~}(#2)~#3}
\def\ap#1#2#3{{ Ann. Phys. }{\bf #1~}(#2)~#3}
\def\ptp#1#2#3{{ Prog. Theor. Phys. }{\bf #1~}(#2)~#3}
\def\plb#1#2#3{{ Phys. Lett. }{\bf#1B~}(#2)~#3}
\def\apjl#1#2#3{{ Astrophys. J. Lett. }{\bf #1~}(#2)~#3}
\def\n#1#2#3{{ Nature }{\bf #1~}(#2)~#3}
\def\apj#1#2#3{{ Astrophys. Journal }{\bf #1~}(#2)~#3}
\def\anj#1#2#3{{ Astron. J. }{\bf #1~}(#2)~#3}
\def\mnras#1#2#3{{ MNRAS }{\bf #1~}(#2)~#3}
\def\grg#1#2#3{{ Gen. Rel. Grav. }{\bf #1~}(#2)~#3}
\def\s#1#2#3{{ Science }{\bf #1~}(19#2)~#3}
\def\baas#1#2#3{{ Bull. Am. Astron. Soc. }{\bf #1~}(#2)~#3}
\def\ibid#1#2#3{{ ibid. }{\bf #1~}(19#2)~#3}
\def\cpc#1#2#3{{ Comput. Phys. Commun. }{\bf #1~}(#2)~#3}
\def\astp#1#2#3{{ Astropart. Phys. }{\bf #1~}(#2)~#3}
\def\epj#1#2#3{{ Eur. Phys. J. }{\bf C#1~}(#2)~#3}

\end{document}